\documentclass[preprint]{vldb}
\usepackage{graphicx}
\usepackage{balance}
\usepackage{subcaption}
\usepackage{xcolor}
\usepackage[level]{datetime}
\usepackage{amsmath}
\usepackage{bm}
\usepackage[font={small}]{caption}
\usepackage{bold-extra}
\usepackage[linesnumbered,ruled]{algorithm2e}
\usepackage{enumitem}
\usepackage{caption}
\usepackage{booktabs}
\usepackage[switch, pagewise]{lineno}
\SetKwProg{Fn}{Function}{}{end}

\newcommand{\projectname}{ArrayBridge}
\newcommand{\papertitle}{\projectname: Interweaving declarative array processing with high-performance computing}

\usepackage{fancyhdr}
\usepackage[breaklinks]{hyperref}
\usepackage[hyphenbreaks]{breakurl}
\hypersetup{pdftitle={\papertitle}}
\hypersetup{pdfauthor={}}
\toappear{}

\begin{document}
\newcommand*{\refname}{Bibliography}

\title{\papertitle}
\newcommand*{\osu}{\textsuperscript{$\bigstar$}}
\newcommand*{\lbnl}{\textsuperscript{\textdagger}}
\newcommand*{\paradigm}{\textsuperscript{\textsection}}
\numberofauthors{1}
\author{
Haoyuan Xing\osu,\hspace{0em}
Sofoklis Floratos\osu,\hspace{0em}
Spyros Blanas\osu,
\vspace{0.2em}
\\
Suren Byna\lbnl,\hspace{0em}
Prabhat\lbnl,\hspace{0em}
Kesheng Wu\lbnl,\hspace{0em}
Paul Brown\paradigm
\vspace{0.8em}
\\
\affaddr{
\hspace{1.5em}
\osu\hspace{0em}
The Ohio State University\hspace{3.25em}
\lbnl\hspace{0em}
Lawrence Berkeley National Laboratory\hspace{2em}
\paradigm\hspace{0em}
Paradigm4, Inc.
}
\vspace{0.3em}
\\
\affaddr{
\{xing.136, floratos.1, blanas.2\}@osu.edu\hspace{3em}
\{sbyna, prabhat, kwu\}@lbl.gov\hspace{2em}
pbrown@paradigm4.com
}}

\maketitle

\begin{abstract}

Scientists are increasingly turning to datacenter-scale computers to produce and analyze massive arrays. Despite decades of database research that extols the virtues of declarative query processing, scientists still write, debug and parallelize imperative HPC kernels even for the most mundane queries. This impedance mismatch has been partly attributed to the cumbersome data loading process; in response, the database community has proposed \emph{in situ} mechanisms to access data in scientific file formats.  Scientists, however, desire more than a passive access method that reads arrays from files. 

This paper describes ArrayBridge, a bi-directional array view mechanism for scientific file formats, that aims to make declarative array manipulations interoperable with imperative file-centric analyses. Our prototype implementation of ArrayBridge uses HDF5 as the underlying array storage library and seamlessly integrates into the SciDB open-source array database system. In addition to fast querying over external array objects, ArrayBridge produces arrays in the HDF5 file format just as easily as it can read from it. ArrayBridge also supports time travel queries from imperative kernels through the unmodified HDF5 API, and automatically deduplicates between array versions for space efficiency. Our extensive performance evaluation in NERSC, a large-scale scientific computing facility, shows that ArrayBridge exhibits statistically indistinguishable performance and I/O scalability to the native SciDB storage engine.

\end{abstract}


\section{Introduction}

\noindent
Scientists are increasingly turning to datacenter-scale computers to
understand phenomena that would otherwise be impossible or 
intractable to approach experimentally.
Domains as diverse as plasma simulation \cite{bowers08},
cosmology \cite{Habib2013}, and climate modeling \cite{cmip5} use
massively parallel simulations that naturally generate many terabytes of
array data for further analysis.
Many of these arrays are stored in scientific file formats like
HDF5~\cite{folk1999hdf5,folk2011overview} and
netCDF~\cite{rew1990netcdf}, which 
have been deployed in production systems for nearly two decades. 
Usage data from petaflop-scale computing facilities
corroborates that HDF5 and netCDF remain among the most popular building
blocks for scientific computing \cite{zhao2014automatic}.

To quickly sift through data in these formats, scientists commonly write,
debug and parallelize custom analysis kernels that manipulate array data in an
imperative fashion---even
for mundane analyses that could be expressed as a succinct
declarative query.
Despite the growing appreciation by scientists for software that makes
low-level manipulations redundant, systems like SciDB~\cite{brown2010}
and TileDB~\cite{tiledb} are not
widely used in large-scale scientific computing facilities.
One challenge is that loading array data imposes onerous time and storage space
requirements.
For SciDB, in particular, we have observed that parallel loading takes hours
even for modestly-sized ($\sim$100 GiB) arrays
and temporarily uses many times more space than the size of the input for data staging.
Array-centric database systems should query HDF5 data in their native file
format, instead of anticipating scientists to convert existing HDF5
datasets into more efficient formats.

Scientists, however, desire more than a passive \emph{in situ} access method that reads
array datasets from a file. 
In particular, 
efficiently serializing arrays in popular file formats is important
to make declarative array manipulations interoperable with 
file-centric processes such as simulation or visualization.
The technical challenge is that 
many array file format libraries have an inherent single-writer,
multiple-readers (SWMR) design constraint~\cite{robinson2013metadata,
hdf52015swmr}.
This poses an inherent scalability bottleneck because a single writing process 
cannot match the I/O capabilities of the parallel file system 
found in modern scientific
computing facilities.

In addition, many imperative codes (such as simulations) are iterative 
and naturally produce 
versioned array objects.
Although time travel queries have been supported in array database systems
like SciDB, 
scientists lack a principled way of accessing past object versions that
are stored in an HDF5 file.
Saving versions as different objects creates bloated files that
duplicate unmodified chunks. The
opportunity is to intercept versioning semantics from array database systems
to discard redundant array chunks, and in turn conserve I/O bandwidth and storage space.
The technical challenge in supporting time travel queries
over array file formats is backwards compatibility, as former versions should remain
accessible from the version-oblivious HDF5 API.

This paper describes ArrayBridge, a bi-directional array view mechanism
that allows a declarative array database system like SciDB to produce, 
access and update versioned array objects directly in scientific file formats. 
ArrayBridge aims for greater interoperability between declarative array
processing and imperative analysis kernels.
Our prototype implementation of ArrayBridge uses
HDF5 as the underlying array storage library and seamlessly
integrates into the SciDB open-source array database system.
The unique features of ArrayBridge are:

\begin{enumerate}[leftmargin=1.25em,itemsep=0em,topsep=0em]

\item
ArrayBridge implements an array-centric \emph{in situ} processing
operator to evaluate AQL/AFL queries directly on external HDF5 objects. 
By interfacing to the SciDB query execution engine, ArrayBridge
does not make scientists wait for hours for a dataset to be loaded and
redimensioned before SciDB answers the first AQL/AFL query.

\item
ArrayBridge efficiently materializes array objects in the HDF5 file
format by writing in parallel through a \emph{virtual view}.
The view partitions the writes into independent streams, which
bypasses the single-writer design constraint
that limits the I/O scalability of file format libraries.
When reading, scientists access a single array object through the view
without any modifications to their applications.

\item
ArrayBridge permits time travel queries from existing applications that
access past versions through the unmodified HDF5 file format API.
Under the hood, ArrayBridge deduplicates identical regions between
object versions to reduce the storage footprint of the HDF5 file.
\end{enumerate}

Our extensive experimental evaluation in NERSC, a large-scale scientific computing
facility, shows that parallel loading of TiB-sized binary array datasets
in SciDB can take more than 7 hours and requires many times more staging
space than the original input.
ArrayBridge completely bypasses the cumbersome data loading
process and 
queries multi-TiB array datasets in the HDF5 format in a few minutes.
ArrayBridge exhibits statistically
indistinguishable I/O scalability and performance to the native SciDB
storage engine. In fact, ArrayBridge processes
data at a similar rate to a hand-tuned, imperative C kernel up to 32
nodes; scaling slows in larger clusters due to SciDB processing bottlenecks.
In addition, ArrayBridge materializes SciDB objects into the HDF5 format
nearly as efficiently as SciDB serializes the database in its
proprietary backup-oriented format.
Finally, ArrayBridge exposes old versions as HDF5 datasets and
transparently eliminates redundancy in versioned objects; this feature is
particularly useful for interoperability with version-oblivious
imperative analyses.

The remainder of the paper is structured as follows.
Section~\ref{sec:background} presents details of the SciDB database
system and the HDF5 format that are relevant for ArrayBridge.
The internals of ArrayBridge are described next:
Section~\ref{sec:system-overview} gives a system overview,
Section~\ref{sec:scan} presents the scan interface, while
Section~\ref{sec:parallel_writing} presents the write interface and the
time travel support.
We then present the experimental evaluation in
Section~\ref{sec:experiments} and summarize the lessons learned
when designing the \projectname\ in Section~\ref{sec:lessons}.
Finally, Section~\ref{sec:relwork} discusses related work and
Section~\ref{sec:conclusions} concludes.


\section{Background}
\label{sec:background}

\subsection{The SciDB database system}
\label{sec:background:scidb}

\noindent
SciDB organizes data into \emph{multi-dimensional arrays}. An array
schema defines the number of dimensions the array has (its rank) and the length of each
dimension.
We use \emph{shape} to collectively refer to the rank and the lengths of
the dimensions of an array.
Every cell in an array contains
one or more attributes (values), similar to a tuple in relational databases.
The order and the type of the attributes in each cell are the same, and they
are defined in the array schema. Array schema metadata is
stored in the SciDB catalog, which is a centralized PostgreSQL database.

An array is stored and
processed in the granularity of \emph{chunks} to accelerate I/O performance.
SciDB uses the \emph{regular
chunking} strategy~\cite{soroush2011arraystore} which partitions a multi-dimensional
array into identically-shaped hyper-rectangles.
SciDB stores a chunk using Run
Length Encoding (RLE); the data is stored as one or more segments,
each of which is a tuple $\langle length, same, data\rangle$, where
$length$ is the number of elements in the segment, $same$ represents
whether the elements in the segment are all the same. If $same$ is true,
$data$ is just the element being repeated; otherwise, $data$ is the raw
data stored in a vector form.

SciDB adopts a shared-nothing model, and partitions the chunks across all
instances in a cluster.
Each instance processes its own data, and only
redistributes data across instances if the query plan explicitly
asks so.
One instance acts as the coordinator. During query processing,
the coordinator parses and optimizes the query,
orchestrates the evaluation of partial query fragments among instances,
and returns the final result to the user.

\subsubsection*{Loading data into SciDB}

A user can load data into SciDB using the \texttt{load()} operator,
which supports loading in
parallel from all instances. Parallel loading
requires a separate file for each instance. 
Because popular formats like CSV may serialize multi-dimensional
arrays in a different order than SciDB, loading from these formats
is a two-step process:
First, the user issues a \texttt{load()} statement to create a
one-dimensional array where the coordinates and the variables are saved
as separate attributes, akin to a relational table. 
This one-dimensional representation needs to be converted
to a multi-dimensional array to
benefit from the expressive power of the array data model.
The user subsequently performs a \emph{redimension} operation
to create a multi-dimensional array from the one-dimensional coordinate attributes.

This two-step loading process is inefficient both in terms of
computation and storage. Apart from the computation cost of
converting and rearranging data, extra staging space is also needed in
the process.
Loading an $r$-dimensional array with $a$ attributes 
requires staging space for $r+a$ attributes to stage the
one-dimensional output of the \texttt{load()} operation. The
following \emph{redimension} operation then compacts this array
to store $a$ attributes only. Loading massive
multi-dimensional arrays with a single attribute is particularly
inefficient: for example, a 5-dimensional, single-variable array that is
1 TiB in size would temporarily require 6 TiB of staging space during
loading.

\subsubsection*{Supported file formats}

Plain text formats like comma-separated value (CSV)
files are very versatile and widely used, but these formats are also verbose
and could impose tokenization and parsing overheads.
These disadvantages are exacerbated as the data volume grows.
Binary formats are thus more common in large-scale scientific computing,
as they trade versatility for compactness and performance.
SciDB supports two binary formats for data input and
output: the \texttt{opaque} and \texttt{binary}
formats~\cite{scidbmanual}. 

The \texttt{opaque} format simply copies each RLE-encoded chunk along with its
metadata information directly onto disk. This format is mainly used
for database backups.
An application that reads or produces data in this format
needs to understand how SciDB organizes the chunk data
internally. Hence, the \texttt{opaque} format is not suitable
for scientific applications that desire interoperability.

The \texttt{binary} format concatenates the binary representation of
every attribute for each cell, and serializes cells in row-major order.
This process is time-consuming as it requires extracting, converting and
copying cells from their native RLE representation into the
\texttt{binary} format.

\subsection{The HDF5 file format}
\label{sec:background:hdf5}

\noindent
The HDF5 format~\cite{folk1999hdf5} is a prominent scientific data format.
Data in HDF5 files is organized using two key
objects: datasets and groups. A \emph{dataset} is a
multi-dimensional array containing the same type of elements.
The data in an HDF5 dataset can be split into multiple
\emph{chunks}, each of which stored in a separate contiguous block in the file.
HDF5 uses the same chunking method as SciDB.
Similar to
directories in a file system, \emph{groups} organize data objects
such as datasets and other groups into a hierarchical structure.

An HDF5 array is always logically dense. However, applications can
register a well-defined value as the \emph{fill value} of a dataset.
If no chunk has
been created, HDF5 will return the fill value on access. This saves
space for arrays with contiguous empty regions, such as a triangular
matrix.

\subsubsection*{The virtual dataset feature}

A recent feature of HDF5 is
\emph{virtual
dataset} support \cite{rfchdf5vd,rees2015developing} that combines data from multiple
source datasets. A virtual dataset can be accessed as an ordinary
dataset, but does not store actual
data. A virtual dataset defines a list of
mappings $m_1,m_2,m_3,...,m_n$, that describe where the actual
data is stored. A mapping is represented as a tuple $<d, src, dst>$; a
source dataset $d$, where the actual data is stored; a source selection
$src$ marks the elements in $d$ that are parts of the virtual
dataset; and a target selection $dst$ marks the logical positions
of the source elements in the virtual dataset.

When a program
reads or writes a region in a virtual dataset, the HDF5 library
traverses the mapping list to find all the mappings that intersect with the
queried region, and then propagates the reading or writing operation to
the corresponding source datasets sequentially. 
As of version 1.10, the HDF5
library does not support removing items from the mapping list,
therefore, the only way to modify a virtual dataset is to recreate the
list from scratch.


\section{System overview}
\label{sec:system-overview}

\begin{figure}[t]
\centering
\includegraphics[width=\columnwidth]{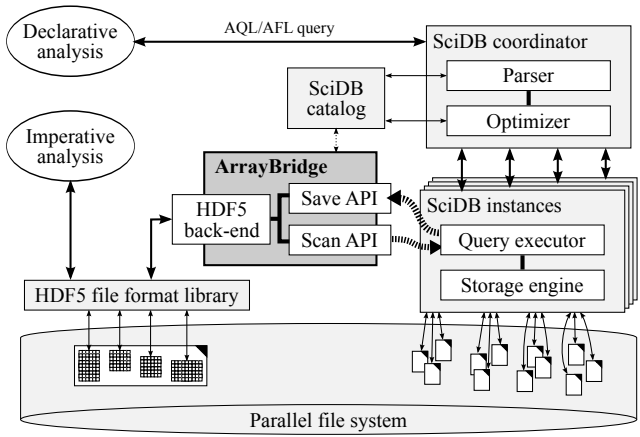}
\caption{
Through ArrayBridge, scientists can intermingle declarative and
imperative analyses over the same dataset. The prototype implementation
of ArrayBridge is integrated with SciDB and uses HDF5 
for data storage.
}
\label{fig:arraybridge}
\vspace{-1em}
\end{figure}

\noindent
ArrayBridge allows imperative manipulations and declarative queries to be issued
against the same array object on disk. An overview of the system is shown in
Figure~\ref{fig:arraybridge}. ArrayBridge exposes two 
interfaces: a read interface which scans an external array and a
write interface that saves into an external array.
Our prototype implementation of ArrayBridge integrates with the SciDB
query engine
and uses the HDF5 library as its storage back-end.

\label{sec:scan:create}

Users query HDF5 data using AQL/AFL queries.
We refer to an HDF5-resident array as an \emph{external} SciDB array,
to distinguish it from the \emph{native} SciDB arrays.  Whenever an
external array is queried, control is routed to ArrayBridge which
returns the underlying HDF5 data.
The new operator \texttt{create\_array\_hdf5()} declares an external
array.
The user specifies the name and the schema of the array, as
well as the HDF5 file and the dataset name, as such:

\begin{verbatim}
create_array_hdf5(array1, <val1:double>
        [i=0:999,100,0], "data1.hdf5:val1");
\end{verbatim}

This statement
creates a 1000-element array with one double attribute, stored in HDF5
dataset \texttt{val1} in the file \texttt{data1.hdf5}.
This populates metadata about the array schema in
the SciDB catalog (a PostgreSQL database).

One SciDB array can contain multiple attributes (columns), whereas
data in HDF5 is organized into single-attribute datasets. Hence,
an external array can contain more than one HDF5 datasets
with the same shape, each represented as one attribute
of the array. Users create multiple attributes in an external array
by listing all the attributes and the corresponding datasets
in the \texttt{create\_array\_hdf5()} statement.


\section{Reading external array objects}
\label{sec:scan}
\label{sec:scan:readext}

\noindent
This section presents how ArrayBridge reads HDF5 data and exposes
them to SciDB. 
After an external array is created, a user retrieves the contents of the
external array in an AFL query using the \texttt{scan\_hdf5()} operator.
Depending on the query,
the \texttt{scan\_hdf5()} operator either reads
the entire array or selectively retrieves specific chunks of the array.
This new operator has the same semantics as the 
\texttt{scan()} operator that accesses native SciDB arrays.

\subsection{Design of the Scan operator}

\noindent
At the planning stage, SciDB parses an AFL query
and generates a query plan for optimization.
The query plan is a tree of array operators. 
During the optimization phase, 
each operator reports the
schema and the estimated size of its output array. 
(\projectname\ uses the array metadata in the SciDB catalog for this information.)
Each operator in SciDB exposes a chunked-based iterator
interface that returns one array chunk of its output upon request.

The interface to the \texttt{scan\_hdf5()} operator consists of three
functions: $Start(obj, attr)$, $Next()$ and
$SetPosition(pos)$. 
The operator is initialized by calling $Start()$ with two parameters that
indicate the requested array object $obj$ and the requested
attribute $attr$ in that object. After the operator is initialized, calling $Next()$
repeatedly iterates through the array and returns the next chunk
assigned to this instance.
If the query needs to retrieve specific array chunks, the parent
operator will intersperse calls to the $SetPosition()$ function in
between calls to $Next()$. $SetPosition()$ accepts a coordinate $pos$ as
its parameter for random accesses to the external array object.
If the chunk at $pos$ is assigned to this instance, $SetPosition(pos)$
returns true; the next invocation of the $Next()$ function
will retrieve the chunk that contains the element at $pos$.
$SetPosition(pos)$ returns false if the chunk at $pos$ is not assigned
to this instance.

Given that ArrayBridge processes data concurrently on multiple nodes, 
one aspect of the design is how to assign the chunks of an array to
different SciDB instances. The native SciDB storage engine assigns
chunks to instances when \texttt{create\_array()} is called, and
stores this assignment in the SciDB catalog.
This design choice is limiting, however, because imperative applications
can change the shape of external array objects through direct calls to the
HDF5 API and leave SciDB with stale metadata.
ArrayBridge has significant more leeway on
when to map chunks to instances because external files on a parallel
file system are visible to all instances.
ArrayBridge thus assigns chunks to instances at query time. 
Assigning chunks to instances at query time
mitigates load skew and presents an opportunity to update stale metadata
to their correct values.

\begin{figure}[t]
\centering
\includegraphics[width=\columnwidth]{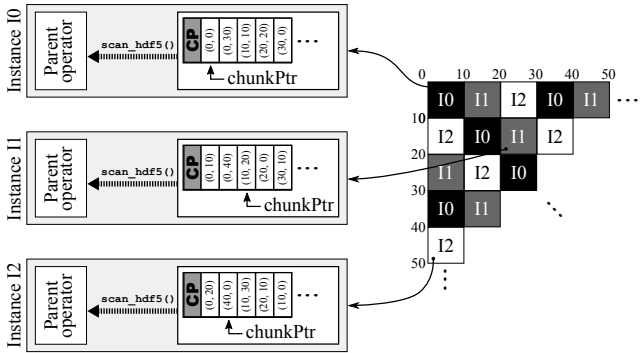}
\vspace*{-1.5em}
\caption{
SciDB processes data directly from an external HDF5 array using the
\texttt{scan\_hdf5()} operator of \projectname.
}
\label{fig:scan:access_method}
\vspace*{-1em}
\end{figure}

\subsection{Implementation of the Scan Operator}

\begin{algorithm}[t]
\caption{The \texttt{scan\_hdf5()} operator.}
\label{algo:scan_hdf5}
	\Fn{Start(Array $obj$, Attribute $attr$)}{
		$(f, d) \leftarrow$ lookup $(obj, attr)$ in the SciDB catalog

		open the HDF5 file $f$ and the HDF5 dataset $d$

		empty $CP$

		\ForEach{\textrm{chunk} $p_i$ \textrm{\textbf{in} dataset} $d$}
		{
			\textbf{if} $\mu(p_i) =$ this instance \textbf{then} add $p_i$ into $CP$
		}
		
		$chunkPtr \leftarrow $ first element in $CP$

	}

	\Fn{Next($ $)}{
			$c \leftarrow $ a new RLE chunk with unique elements

			$r \leftarrow $ region at $chunkPtr$ to read from

			call \texttt{H5Dread}($d, r, c$) to read data into $c$

			advance $chunkPtr$

			\Return $c$
	}
	
	\Fn{SetPosition(Coordinate $pos$)}{
		$\widehat{p} \leftarrow $ the coordinates of the chunk that contains $pos$

		\If{$\widehat{p} \in CP$}{
			$chunkPtr \leftarrow \widehat{p}$

			\Return True
		}
		\Return False
	}
\end{algorithm}

\noindent
We now describe the implementation details of the \texttt{scan\_hdf5()}
operator. The pseudocode is shown in Algorithm \ref{algo:scan_hdf5}.

\vspace{0.25em}
\textbf{The Start() method:} 
Before the scan operation commences, 
all instances first determine how
to partition the chunks to parallelize the scan operation.
We identify each chunk by its coordinates, and use a mapping
function $\mu()$ to abstract different chunk assignment algorithms
\cite{rusu2013survey}. We use the round-robin assignment in our implementation.

Each instance creates an ordered array $CP$ in memory that stores all the
chunks that the mapping function has assigned to this instance. 
When $Start(obj, attr)$ is called, 
the catalog is consulted to translate an $(obj, attr)$ reference into
the appropriate HDF5 file name $f$ and the 
dataset $d$.
$Start()$ then opens file $f$, 
reads the array shape information for dataset $d$,
and iterates over all chunk positions. 
Instance $i$ adds chunk $c$ to its local $CP$ if and only if 
$\mu(c) = i$.
Finally, the chunk read pointer $chunkPtr$ is initialized to
the first chunk.

\vspace{0.25em}
\textbf{The Next() method:} 
$Next()$ reads the chunk at the coordinates pointed by $chunkPtr$
from the HDF5 dataset, and advances $chunkPtr$.
Reading data from an HDF5 file into memory is done by
calling \texttt{H5DRead()} and passing the source dataset $d$, the source 
region $r$, and the destination buffer $c$ to store the in-memory data. 
The HDF5 library uses the C array representation for in-memory data.

The technical challenge is to efficiently convert HDF5 array data from their C
representation into the RLE format that the SciDB operators use.
We discovered that converting the data into the RLE representation 
is a serious performance hit for the dense arrays that HDF5 is
well-optimized for. 
We instead opted to masquerade a dense HDF5 chunk as an
RLE-compressed chunk $c$ where each element is unique, and 
request \texttt{H5DRead()} to place the data directly into $c$.
This avoids redundant data copying and format conversions.

\vspace{0.25em}
\textbf{The SetPosition() method:} 
$SetPosition(pos)$ first computes the coordinates $\widehat{p}$ of the
chunk that contains $pos$, and performs binary search in the $CP$ array to find
$\widehat{p}$. If $\widehat{p}$ is found,
$SetPosition$ changes $chunkPtr$
to $\widehat{p}$ and returns true, otherwise it returns false.


\section{Saving array objects in files}
\label{sec:parallel_writing}

\noindent
Many post-processing and visualization workflows are conducted over 
file-centric APIs. It is thus important to
efficiently serialize arrays into popular scientific data formats such as HDF5.
This section presents how \projectname\ materializes
SciDB objects into external HDF5 files. 
The two key features that \projectname\ supports are (1) parallel writing
that bypasses the single-writer limitation of HDF5, and (2) time travel so
that imperative kernels can access previous versions of a dataset over
the HDF5 API without modifying the code. We describe each in turn.

\subsection{Balancing efficiency and interoperability}

\begin{figure}[t]
\centering
\includegraphics[width=\columnwidth]{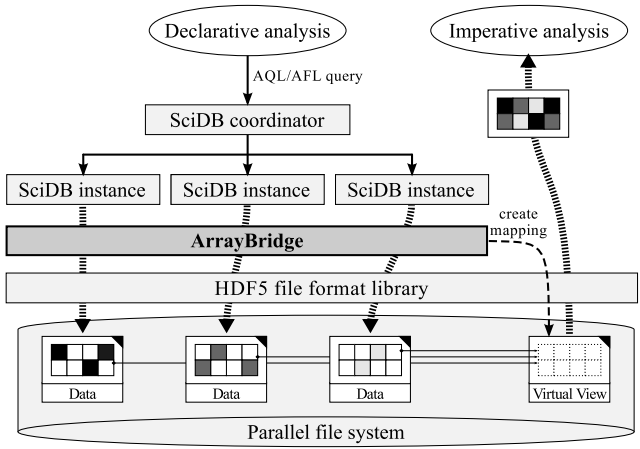}
\caption{
The virtual view mechanism bypasses the single-writer limitation of
the HDF5 file format by directing write streams from different SciDB
instances into separate objects. Existing imperative analyses 
access a single object through the view.
}
\label{fig:virtualview}
\vspace*{-1em}
\end{figure}

\noindent
Although an HDF5 file can be read by multiple SciDB instances
concurrently, only one SciDB instance can be writing to a single
HDF5 file at a time.
This fundamental Single-Writer, Multiple-Readers (SWMR) design
constraint of HDF5 is a roadblock to parallel writing and it prevents SciDB
from taking advantage of the highly concurrent I/O subsystems found in
modern scientific computing facilities.

Serializing array data into the HDF5 format exposes a dilemma between
\emph{writing efficiency} and \emph{interoperability}. To write data in
parallel, SciDB must produce multiple HDF5 files. Splitting a single
dataset into multiple files, however, makes analysis more cumbersome.
Conversely, storing the entire dataset in a single file makes it
straightforward to manage and analyze the HDF5 dataset from existing
applications, but it limits the write throughput to the
I/O capacity of one SciDB instance.

SciDB exposes this trade-off between efficiency and interoperability to the user. 
The \texttt{save()} operator provides two
separate writing modes. In the \emph{Serial} mode, the data is
shuffled to the SciDB coordinator which writes all the data to a
single file. This way, processing and managing the data is straightforward, 
but the writing throughput is limited to the I/O capacity of one
instance. In the \emph{Partitioned} mode, each instance
writes the data it stores to a separate file. This way, the writing is
parallelized, but the user needs to deal with the complexity of
maintaining partition information and adapting existing tools to read
partitioned files.

\projectname\ supports a third writing mode, the \emph{Virtual View} mode,
to avoid the artificial dilemma between sacrificing writing
efficiency or interoperability. 
The \emph{Virtual View} mode writes the data on each instance to
separate HDF5 files, and then creates a virtual dataset that logically
combines these datasets into one object. 
The \emph{Virtual View} mode utilizes the virtual dataset feature
of the HDF5 format, which presents the data in several HDF5
datasets as a single dataset to the application accessing it. The
benefit of the \emph{Virtual View} mode is that it combines the
efficiency of parallel writing into separate files with the
interoperability of producing a single file for the small upfront
cost of creating the virtual dataset.

\subsection{Serializing SciDB arrays into HDF5}

\noindent
\projectname\ extends the SciDB \texttt{save()} operator to support the HDF5
format. This allows a user to write data back to external HDF5 files.
The \emph{Serial} and the \emph{Partitioned}
writing modes adopt the SciDB semantics: In the
\emph{Serial} mode, data is shuffled to the coordinator, and the coordinator
writes the data into an HDF5 file. In the \emph{Partitioned} mode, each instance
creates a different file and saves the data assigned to it to a dataset with
the same shape as the original array. The
chunks that have not been assigned to the local instance are empty: HDF5
logically fills them with a \emph{fill value} but does
not store any additional data.

The \emph{Virtual View} mode
starts by having every instance 
write its chunks to a separate file, just like in the \emph{Partitioned} mode. In
addition to writing the chunks, each instance also maintains two 
regions: one is the source region in the local file ($src$), the other
is the target region in the virtual dataset ($dst$).
After the chunks are written, the \emph{Virtual View} mode needs to
create the virtual dataset.

\projectname\ implements two methods to
create the virtual dataset.
The first method is \emph{parallel} mapping. 
Each instance appends its mappings
into the virtual dataset, using the $src$ and $dst$ objects it created
during the writing process. The virtual dataset needs to be recreated, as
it cannot be
updated directly. Thus, each instance needs to read the
mappings in the virtual dataset, append its own mapping to it, and
recreate the dataset using the new mapping list. Because only
one instance can update the virtual dataset at a time, the
parallel mapping technique uses
file locking to ensure mutual exclusion.
This crude synchronization method allows each instance to update its
own mapping without waiting for other instances to exchange mapping data.
However, because each update recreates the dataset,
$O(n^2)$ mappings will be written for a cluster with $n$ SciDB
instances.

The second method of creating the virtual dataset is
\emph{coordinator} mapping. Each instance transmits the
$src$ and $dst$ regions to the coordinator, which concatenates the
per-instance mapping lists and creates the
virtual dataset. This requires all SciDB instances to
synchronize on a barrier and wait for
the coordinator to create the virtual dataset. The coordinator
mapping technique writes $O(n)$ mappings for a
cluster with $n$ SciDB instances.
We evaluate the
performance trade-offs of these two techniques in
Section~\ref{sec:exp:save}.

\subsection{Backward-compatible time travel in HDF5}

\noindent
An oft-requested feature is the ability to query past versions of a
dataset to understand how a dataset evolved or to explain the
processing steps to the final result.
This subsection introduces the time travel capability in \projectname\
that transparently deduplicates regions that are identical between
dataset versions that are stored in the HDF5 file. This reduces the
storage footprint of versioned datasets and is interoperable with
imperative applications that access former versions through the
existing HDF5 API.

Because analyses predominantly access the latest version of a dataset,
\projectname\ always fully materializes the latest version to minimize
reconstruction costs.
Past versions are stored under a separate group in the HDF5 object
hierarchy. 
All versions can be accessed as ordinary HDF5 datasets by 
applications via the HDF5 API.
A user produces a versioned dataset by passing a parameter in the
\texttt{save()} operator; \projectname\ accesses old versions as ordinary
HDF5 datasets via an explicit call to the \texttt{scan\_hdf5()} operator.
\projectname\ supports two techniques for saving a versioned dataset: Full Copy
and Chunk Mosaic.

\textbf{Full Copy:}
The Full Copy technique incurs the cost of materializing every past version.
For example, if a query calls \texttt{save()} to update the dataset
\texttt{speed} to version \texttt{V2}, \projectname\ renames dataset
\texttt{speed} as \texttt{PreviousVersions/V1}. After this metadata
operation is completed, the \texttt{save()} operator creates
a new dataset named \texttt{speed} that stores the latest
version \texttt{V2}.

\begin{figure}
\centering
\includegraphics[height=1.3in]{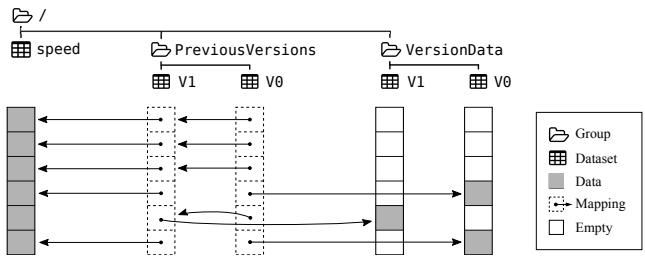}
\caption{
An example of the Chunk Mosaic technique. The latest
version is the fully-materialized dataset \texttt{speed}. 
Old versions are accessed through
virtual datasets in the \texttt{PreviousVersions/} group.
}
\label{fig:save:version}
\vspace*{-1em}
\end{figure}

\textbf{Chunk Mosaic:} The Chunk Mosaic technique only stores the 
chunks that are updated and creates a virtual dataset to ``stitch
together'' the past version.
An example of the Chunk Mosaic technique is illustrated in Figure
\ref{fig:save:version}. 

The Chunk Mosaic technique proceeds in two steps. The first step is 
creating an HDF5 dataset that stores the previous versions of the array
chunks that are updated. The unmodified chunks are empty.
(Recall that the HDF5 library does not store empty chunks).
This dataset has the same shape as the original dataset, and
is hidden away in an HDF5 group labeled \texttt{VersionData/}.
For example, in Figure~\ref{fig:save:version}, the HDF5 dataset
\texttt{VersionData/V0} only stores the chunks of version \texttt{V0} of the
\texttt{speed} dataset that were updated in version \texttt{V1}.
Because SciDB does not convey which chunks were updated 
to the \texttt{save()} operator, 
our current implementation discovers which chunks
changed by comparing the chunk that is being saved with the chunk that
already exists in the latest version of the dataset.

The second step of the ChunkMosaic technique creates a virtual dataset 
under the \texttt{Pre\-viousVersions/} group in the HDF5 file. This
dataset combines
the unmodified chunks of the latest version and the updated
chunks in the \texttt{VersionData/} group into a single view.
This virtual view will be accessed by applications to retrieve
this old version.
In the example shown in Figure~\ref{fig:save:version}, suppose that the
update creates version \texttt{V2} which is stored in dataset
\texttt{speed}. The Chunk Mosaic technique first
creates the virtual dataset \texttt{PreviousVersions/V1}. 
If a chunk
was modified, the virtual dataset chunk is mapped to 
the \texttt{VersionData/V1} dataset that contains the original chunk
in the previous version;
otherwise the chunk is unmodified and is mapped to its latest version
at \texttt{speed}.
Finally, the mappings in the dataset
\texttt{PreviousVersions/V0} that point to \texttt{speed} are modified
to point to \texttt{PreviousVer\-sions/V1}.
The Chunk Mosaic technique produces a
series of chained virtual datasets that can reconstruct any previous
version by either referencing the latest fully-materialized version or
a former version of a chunk in the \texttt{VersionData/} group.


\section{Experimental Evaluation}
\label{sec:experiments}

In this section we experimentally evaluate the 
scan, save and time travel
functionalities of ArrayBridge. We consider the following
questions: 

\begin{itemize}[leftmargin=1.25em,itemsep=0em,topsep=0em]
\item
(\S\ref{sec:exp:scan}) 
How does the performance of \projectname\ compare with 
native SciDB when reading data from a parallel file system in a
typical HPC environment? How much time and space does loading take?

\item
(\S\ref{sec:exp:vpic})
How does \projectname\ scale when processing a real multi-TiB simulation
dataset that would be infeasible to load in SciDB?

\item
(\S\ref{sec:exp:save})
How efficient is the save mechanism of ArrayBridge, and how does it
scale? How effective is the time travel mechanism in deduplicating
versioned array objects?
\end{itemize}

\subsection{Configuration and methodology}
\label{sec:exp:conf}

We evaluate our implementation on the Edison computer of the National Energy Research
Scientific Computing (NERSC) facility. Edison is a Cray XC30 supercomputer with
5,586 compute nodes. Each node
has two 12-core 2.4Ghz Intel ``Ivy
Bridge'' processors and 64 GiB memory.
File storage is provided via 
the Lustre~\cite{lustre} parallel file system. Lustre distributes
the content of a file across multiple I/O servers, called Object Storage Targets (OST), to provide
I/O concurrency for highly-parallel applications. Each file is divided into
a user-defined \emph{stripe size}, and then these stripes
are distributed to the requested OSTs in a round-robin fashion. The number
of OSTs a file is distributed to is called the \emph{stripe count}. 
The file system we use has a total of 248 Lustre Object
Storage Targets (OSTs) and more than 30 PiB of total storage. The
reported peak I/O throughput exceeds 700 GiB/sec.

We evaluate \projectname\ on the SciDB 15.12 Community
release. According to the SciDB guidelines~\cite{scidbhardware}, we
configure 8 instances per node and a 16 MiB chunk size. The
HDF5 datasets are chunked using the same chunk size (16 MiB).
Unless we specify otherwise, we 
follow the NERSC-recommended I/O configuration of
striping each file to 72 Lustre OSTs
with a 1 MiB Lustre stripe size.

One challenge in reporting performance results from large-scale HPC
facilities such as NERSC is that the computers are always busy with multiple
concurrent jobs.
This brings high variability in I/O performance.
We have sometimes observed variance that exceeds 100\%
for the same experiment. Many of our results thus report performance in
a box plot to convey the effect of variance.
We draw a thick horizontal band at the median
value, while the bottom and top of the box show the first (25\%) and
third (75\%)
quartiles. The whiskers are drawn at the lowest and highest datum within
$1.5\times$ of the inter-quantile range (IQR).

Another evaluation challenge is that there is no mechanism to evict
cached objects from the parallel file system, because caching happens in
multiple layers of the I/O path to cold storage.
We thus revert to yielding the computer to other jobs to clear the
parallel file system cache.
In our experiments, we wait for at
least one hour before reading the same data again to ameliorate any
caching effects.

\subsection{Scan}

\label{sec:exp:scan}
This subsection evaluates the efficiency of analyzing HDF5 file
directly using \projectname\ through queries on synthetic
datasets.

\begin{figure}[t]
	\centering
	\includegraphics[width=\columnwidth]{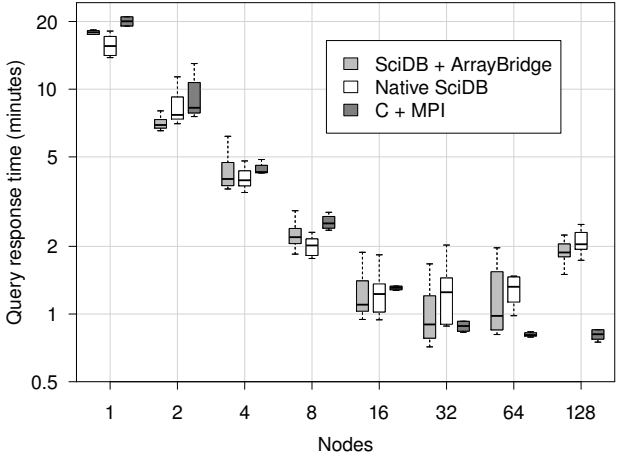}
	\caption{Time to aggregate a 1.5 TiB dataset with different cluster sizes.}
	\label{fig:exp:scan:access_method}
	\vspace{-1em}
\end{figure}

\begin{figure}[t]
\begin{minipage}[c]{0.49\columnwidth}
	\centering
	\includegraphics[width=\textwidth]{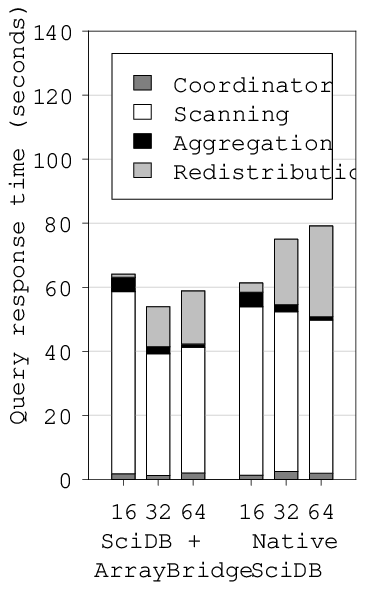}
  \begin{minipage}[c]{0.9\textwidth}
	\captionof{figure}{Time breakdown to aggregate the entire 1.5 TiB dataset.}
	\label{fig:exp:scan:breakdown}
  \end{minipage}
\end{minipage}
\hspace{0em}
\begin{minipage}[c]{0.49\columnwidth}
	\includegraphics[width=\textwidth]{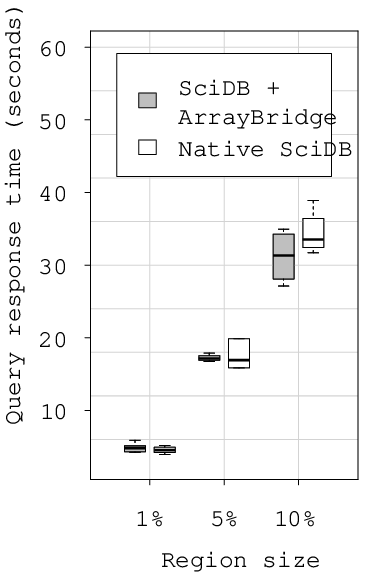}
  \begin{minipage}[c]{0.9\textwidth}
	\captionof{figure}{Response time to aggregate
	contiguous regions in the 1.5 TiB dataset.}
	\label{fig:exp:scan:block_selection}
  \end{minipage}
\end{minipage}
\vspace{-2em}
\end{figure}

\subsubsection*{Access method comparison}
\label{sec:exp:scan:explain}
The first question is how the full
scan performance of \projectname\ compares with the native SciDB storage
engine. 
We choose aggregation as a representative query for the full-scan pattern,
and use an
one-dimensional dataset of 192 billion double numbers (approximately 1.5
TiB data). 
For comparison purposes, we also implemented an imperative aggregation 
kernel in C that is parallelized through MPI and processes
data directly through the HDF5 API.
We vary the number of nodes from 1 to 128, and report the
query response time in Figure \ref{fig:exp:scan:access_method}.

As the result indicates, the query response time of
\projectname\ is as good as native SciDB: the 
performance of \projectname\
is statistically indistinguishable from the performance of the native
SciDB engine.
All implementations scale near perfectly from 1 to 32 nodes.
\projectname\ achieves its best performance with 32 nodes,
aggregating 1.5 TiB of data at a rate of $\sim$ 28 GiB/s.
However, while scaling beyond 32 nodes, the response time of
the imperative C kernel flattens out, but the response time (and
variance) for both \projectname\ and native SciDB increase. 

Figure
\ref{fig:exp:scan:breakdown}
reports the time breakdown when using 16, 32, and 64 nodes 
to shed light on why query response time increases in SciDB.
We decompose the query response time into
four parts: \emph{coordinator time}, the time coordinator uses
to parse and optimize the query, broadcast the query plan, and send the
result to client; \emph{scanning time}, the average time used
to read chunk data from HDF5 library or SciDB storage engine by each
instance; \emph{aggregation time}, the average time
to aggregate data in local chunks at every instance; and
\emph{redistribution time}, the time to redistribute the local
aggregation results to the coordinator.

In our experiment, the scanning time for both the \projectname\ and native
SciDB flattens beyond 32 nodes. This mimics the I/O behavior of the
imperative C kernel (see Figure~\ref{fig:exp:scan:access_method}).
The aggregation
time does scale perfectly, but it only accounts for a small portion
of the query response time, whereas the coordinator time stays
constant and only occupies a
small portion of the response time. 
The culprit for the increased query response time is the redistribution
cost, which significantly increases with more than 32 nodes. 
This also explains why the imperative kernel outperforms
\projectname\ and native SciDB with more than 64 nodes. 
Profiling the imperative kernel revealed that it takes almost the same time to scan the HDF5
file and aggregate locally as for \projectname,\ but it takes far more
time to redistribute partial aggregates in SciDB than to aggregate the
result with MPI. 

Of course, not all queries will access the entire dataset.
To evaluate the effect of implicit indexing in SciDB,
we also run a block selection query that randomly aggregates one contiguous region of the
1.5 TiB dataset using 8 nodes. We vary
the size of the selected region from 1\% to 10\% of the dataset. The result is shown in Figure
\ref{fig:exp:scan:block_selection}. The response time of ArrayBridge and
native SciDB is statistically indistinguishable regardless of the selected region
size. This shows that the SciDB chunk index
does not bring a performance advantage over scanning a chunked HDF5 file
directly even when only retrieving a region of a dense array.

\begin{figure*}
  \captionsetup[subfigure]{justification=centering}
	\begin{subfigure}[b]{2.4in}
		\centering
		\includegraphics[width=\textwidth]{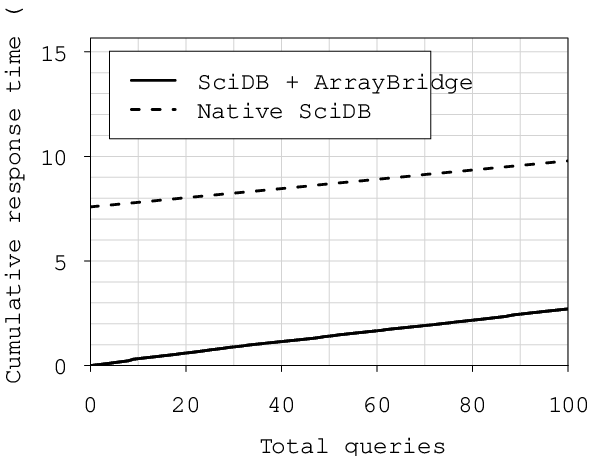}
\caption{Time to aggregate a 1 TiB dataset\\that does not fit in memory.}
		\label{fig:exp:scan:load1Tnersc}
	\end{subfigure}
	\begin{subfigure}[b]{2.4in}
		\centering
		\includegraphics[width=\textwidth]{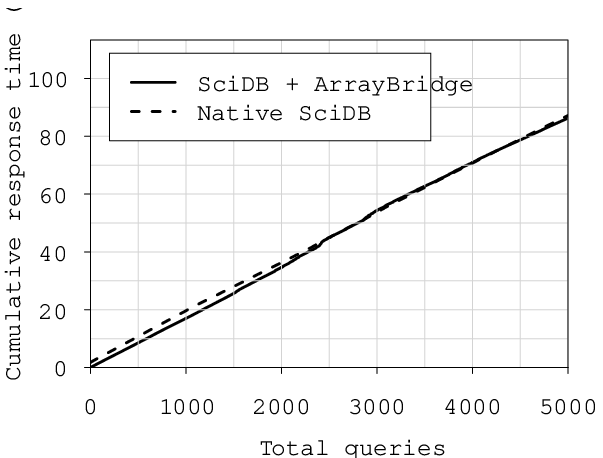}
\caption{Time to aggregate a 16 GiB dataset\\that fits in the SciDB buffer pool.}
		\label{fig:exp:scan:load16Gnersc}
	\end{subfigure}
	\begin{subfigure}[b]{2.4in}
		\centering
		\includegraphics[width=\textwidth]{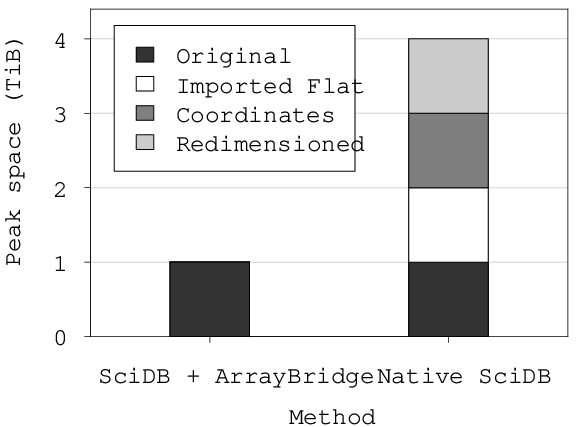}
\caption{Peak storage requirement when\\loading a 1 TiB dataset.}
		\label{fig:exp:scan:loadspace1Tnersc}
	\end{subfigure}
	\vspace{-1em}
	\label{fig:exp:scan:loadnersc}
\caption{Cumulative query response time and peak space consumption	to load and complete multiple queries.}
\end{figure*}

\subsubsection*{Cumulative query response time with loading}

Another question is
how much extra time and space is needed for loading the data,
and whether this upfront time investment accelerates subsequent queries
to the same dataset.
We evaluate these questions by repeatedly aggregating the same
dataset using both \projectname\ and native SciDB.
For \projectname,\ we query the HDF5 dataset directly; for native SciDB,
we load the binary dataset
in parallel and redimension it into an one-dimensional array.
We fix the cluster size to 8 nodes, and use two
synthetic datasets: the first is only 16 GiB and comfortably fits in
the local buffer pool (chunk cache) of each SciDB instance; the
other is 1 TiB which is nearly twice as big as the local memory
in each node.

Figures \ref{fig:exp:scan:load1Tnersc} and
\ref{fig:exp:scan:load16Gnersc}
report the cumulative query
response time of loading and evaluating the first $n$ queries, $n = 1, 2,
3, ...$. The value at $n = 0$ represents the dataset loading time.
For the 1 TiB dataset, 
it takes more than 7.5 hours for the native
SciDB to load the data, redimension it, and return the first
query result. 
(The redimension operation alone takes a little more than 7 hours.)
In comparison, it only takes \projectname\ about 1.5 minutes
to answer the first query!
\projectname\ provides the answer $300\times$ faster
by skipping the time-consuming loading process.

Loading the data into SciDB does not significantly accelerate
subsequent queries, either. 
Surprisingly, this holds even for the 16 GiB dataset which
fits entirely in the buffer pool.
Recall that  the \texttt{scan\_hdf5()} operator issues I/O requests to
the HDF5 file on every call,
unlike native SciDB that can redirect I/O requests to the buffer pool.
We conclude that caching and prefetching by the Lustre
parallel filesystem can be as effective as 
the native SciDB buffer pool in curtailing redundant I/O to cold storage
for full scan access pattern.

Another significant advantage of querying HDF5 data directly in its
native format is reducing
the space overhead of loading.  Figure
\ref{fig:exp:scan:loadspace1Tnersc} breaks down the peak space consumption
during the loading process. As described in Section
\ref{sec:background},
the steps of importing an one-dimensional array into
SciDB includes importing the original data into a flat
SciDB array with coordinate information, and then convert it to the
redimensioned SciDB format. For a one-dimensional array, this occupies
$3\times$ more space than the original data. Datasets of higher
dimensionality would need even more space, as the size of the coordinate data
increases linearly with the dimensionality of the dataset.

\subsection{Analyzing particle-in-cell (PIC) simulation data with \projectname}
\label{sec:exp:vpic}

\begin{figure}[t]
\centering
\includegraphics[width=\columnwidth]{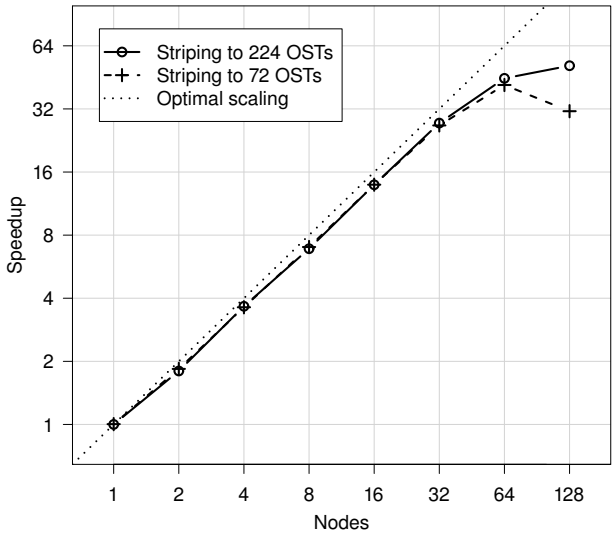}
\caption{Speedup when analyzing 4.3 TiB of post-processed 
Particle-In-Cell simulation data with ArrayBridge. Striping to more Lustre
OSTs better utilizes the parallel file system at scale.}
\label{fig:exp:vpic}
\end{figure}

\noindent
We now evaluate the performance of ArrayBridge when
analyzing a scientific dataset that would be infeasible to
load into SciDB due to its massive size. 
An opportunity for symbiosis
between declarative and imperative analyses occurs during post-processing
of large-scale simulation data.
The dataset is from the plasma physics domain. The 
scientific application 
uses
VPIC \cite{bowers08},
a general purpose particle-in-cell simulation code, for modeling
kinetic plasmas with billions of particles. 
The simulated phenomenon of interest occurs when
an intense femtosecond laser pulse is focused on a solid target to
create a dense plasma that acts as a so-called ``plasma mirror'' and
specularly reflects incident light \cite{Gonsalves:2015:GPS,
Thaury:2007:PMU}. This generates oscillations
of the plasma mirror surface, which in turn deforms the reflected field
by the Doppler effect and produces intense attosecond pulses.
Studying the Doppler harmonic radiation spectrum is challenging, as it
requires simulating and analyzing a 3D field with very high resolution.  
For this problem, one
simulation step produces about 50 TiB of data.

As in many other large simulations,
the current practice is to perform analyses in phases after the
simulation result has been produced. The first phase 
discards low-energy particles of little interest and reduces the dataset
to a 4.3 TiB array in the HDF5 format for
further analysis, visualization, sharing and archival storage.
This 3D dataset consists of four variables: one variable per component
of the 3D velocity
vector $\vec{v} = (v_x, v_y, v_z)$, and
one variable for the energy $E$. The query intent is to
aggregate the velocity $\|\vec{v}\|=\sqrt{v_x^2+v_y^2+v_z^2}$ and energy
$E$ for high-energy particles ($E>2.0$) over a grid. This mundane
analysis can be succinctly conveyed in a SciDB AFL query that directly
accesses these HDF5 objects via ArrayBridge.

The scalability of this aggregation query over the 4.3 TiB array dataset is shown in
Figure~\ref{fig:exp:vpic}. Processing this dataset using a single node returns the answer in 137 minutes.
We observe near-optimal speedup up to 32 nodes. 
With more nodes, the speedup critically depends on the I/O performance. 
The recommended striping policy of 72 OSTs limits scaling beyond 64 nodes
because it limits the available I/O throughput. Striping to 224 OSTs allows for
higher I/O utilization, which improves performance for 64 and 128 nodes. 
Scaling stops at 128 nodes as the internal query
processing operations in SciDB become the bottleneck (see
Section~\ref{sec:exp:scan:explain} for a detailed discussion).

Overall, ArrayBridge can use 64 nodes to aggregate 4.3 TiB of array data
in under 3 minutes for this AFL query. Loading and redimensioning a
dataset of this volume into a 64-node SciDB cluster would have
taken more than a day.

\subsection{Save}
\label{sec:exp:save}

\begin{figure}[t]
\centering
\begin{minipage}{0.49\columnwidth}
\includegraphics[width=\columnwidth]{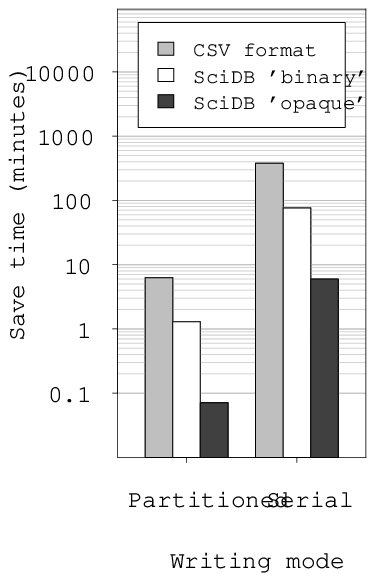}
    \begin{minipage}[c]{0.9\textwidth}
    \vspace{1em}
\captionof{figure}{Time to save a 64 GiB dataset in different formats. }
    \label{fig:exp:save:binary_formats}
    \end{minipage}
\end{minipage}
\begin{minipage}{0.49\columnwidth}
\includegraphics[width=\columnwidth]{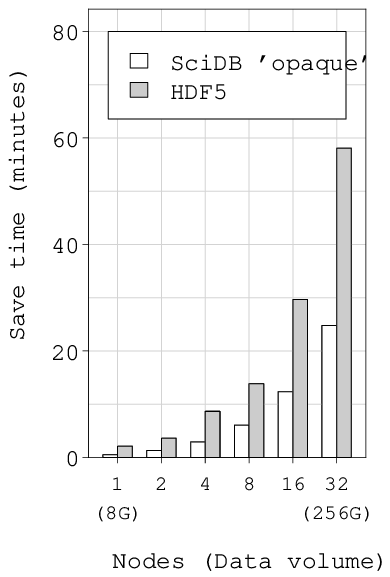}
    \begin{minipage}[c]{0.95\textwidth}
    \vspace{1em}
    \captionof{figure}{Time to save 8 GiB/node in different formats
	with the \emph{Serial} writing mode.}
    \label{fig:exp:save:serial}
    \end{minipage}
\end{minipage}
\vspace{-2em}
\end{figure}

\noindent
SciDB supports multiple formats
for exporting data, including the widely used comma-separated
value (CSV) format, the SciDB `binary' format, and the proprietary 
`opaque' format.
(See Section \ref{sec:background} for details.) 
We evaluate the performance of saving in these
formats by storing a synthetic two-dimensional 64 GiB dataset of
double numbers with 8 nodes, using both the serial
and partitioned writing modes.

Figure \ref{fig:exp:save:binary_formats} reports the median query
response time of each format/mode combination. (Note that the vertical
axis is logarithmic.) As the
figure indicates, the CSV and the SciDB `binary' formats are very
slow to write into. Even with 8
nodes writing CSV data in parallel,
the writing throughput per node is only $\sim$ 20 MiB/s due to the
overhead of converting doubles into a text representation.
The SciDB binary format is almost $5\times$ faster as it avoids text
conversions and merely 
reconstructs the binary on-disk representation from the RLE encoding.
The `opaque' format is
$10\times$ faster than the SciDB
`opaque' format, as it merely dumps the RLE-encoded chunks on disk.
Although the `opaque' format is intended for backups and is not designed
to be interoperable across applications, we report save times for the SciDB
`opaque' format in the experiments that follow as it
better approximates the peak aggregate writing capability of a SciDB cluster.

\begin{figure}[t]
\centering
\includegraphics[width=\columnwidth]{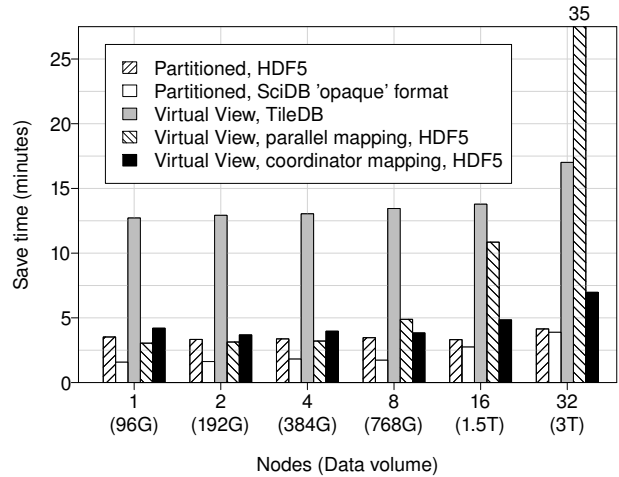}
\caption{Time to save 96 GiB/node using the \emph{Partitioned} and the
\emph{Virtual View} writing modes in HDF5, TileDB and SciDB's
proprietary ``opaque'' formats.}
\label{fig:exp:save:scale}
\vspace{-1em}
\end{figure}

\subsubsection*{Problem of serial writing}
\noindent
We evaluate the performance of the serial writing mode by saving a synthetic
two-dimensional dataset with 8 GiB data per node to disk. (Recall that
the serial writing mode produces a single output file by redirecting all
data to a single SciDB instance.)
We vary the number of nodes from 1 to 8, and report the save time for both the HDF5 format
and the SciDB `opaque'
formats. Figure
\ref{fig:exp:save:serial} shows the median query response time. Although the
computational power and I/O capacity increase with the number of nodes, the
aggregate writing throughput does not increase accordingly as the single
writing instance is the bottleneck; hence, the
response time increases as the data volume increases. Profiling the
result further shows that both shuffling time and writing time double when the
number of nodes doubles. Although materializing arrays
using the serial mode produces a single file, its performance does not scale.

\subsubsection*{Writing in parallel}
We now consider the performance of
writing HDF5 files in parallel using the partitioned and virtual
view
modes. We generate a synthetic two-dimensional
dataset with 96 GiB data per node. Figure 
\ref{fig:exp:save:scale} compares the query response
times of saving this dataset using each format/mode combination.

In terms of scaling, partitioned writing performs well.
Writing HDF5 in partitioned mode scales
perfectly from 1 to 16 nodes, maintaining a writing throughput of
about 450 MiB/s per
node. Writing into the SciDB `opaque' format using partitioned mode
also scales perfectly from 1 to 8 nodes, which records a per-node writing
throughput of about 1 GiB/s. The writing throughput of the
`opaque' format decreases at larger scale and matches the writing
throughput of the HDF5 format for 32 nodes. However, the disadvantage of
the partitioned writing mode is that it produces one file per SciDB
instance. At 32 nodes, the output is contained in 256 files, which can
be cumbersome to manage.

The query response time of the virtual view mode with
local mapping technique is indistinguishable from the partitioned writing
for clusters up to 4 nodes.
However,
the response time goes up while scaling beyond 8 nodes: at 16 nodes, it takes almost
$1.5\times$ longer to write data in the local mapping mode compared with
the partitioned mode; it's almost $10\times$ as
much when using 32 nodes. This is because of the substantial cost of updating
virtual datasets, where each instance reads and re-writes all
prior mappings. Profiling shows that updating the virtual dataset
takes about 90\% of the time for 32 nodes.

The coordinator mapping technique significantly ameliorates this overhead, as the
coordinator collects all the mappings and creates the virtual dataset once.
For 32 nodes, the coordinator mapping technique reduces the virtual
dataset creation time to about 30
seconds, or less than 10\% of the query time. This allows the user to
save a 3 TiB array using virtual view within 7 minutes, and 
preserves the advantage of allowing imperative analysis kernels to access
it as a single dataset.

To evaluate the efficiency of the HDF5 virtual mapping mechanism, we also
compare the performance of saving in the HDF5 format with the
performance of saving in
the TileDB~\cite{tiledb}
format, which
also supports viewing a partitioned array as a single object.
TileDB organizes
arrays into \emph{fragments} that contain ordered elements in a
hyperrectangular array region. Fragments are stored in separate
directories in the underlying file system. This means that different
processes can write to different fragments in parallel without
conflicts. 
Akin to the virtual dataset functionality of HDF5, the TileDB library
scans all the fragments that intersect with the accessed
region, and combines the result at query time.

We implemented a version of the \texttt{save()} operator in
\projectname\ that stores data in TileDB.
The result shows that saving to the TileDB format takes about
$4\sim5\times$ more time than saving in the HDF5 format with virtual view.
Profiling shows that this slowdown is largely caused by the overhead of
creating and closing fragments, which takes nearly 60\% of the
time for a cluster with 16 nodes. In addition, due to
the sheer number of generated directories and files, accessing and
managing
the array dataset is very time-consuming due to the filesystem metadata
management overheads. For example, merely listing the contents of the
directory that stores a 1.5 TiB array in the TileDB format using
\texttt{`ls'} takes more than 30 minutes.

\subsubsection*{Time Travel}
\label{sec:exp:version}
We now evaluate how effectively does the Chunk Mosaic
technique deduplicate versioned datasets during time travel queries when
compared with the Full Copy technique.\footnote{TileDB currently does
not support time travel queries, although this feature is in the roadmap for
a future release.}
We use a synthetic 3.125 GiB dataset of
double numbers with 16 MiB chunk size. To create a new version, we
first identify a range of chunks in the dataset, and randomly change $N$ elements in
these chunks, $N = 1\%$ of the total elements in the array.
The number of chunks being updated is varied from 1\% of total
chunks (2), to 100\% of total chunks (200).

\begin{figure}
\begin{subfigure}[b]{1.6in}
\centering
\includegraphics[width=\columnwidth]{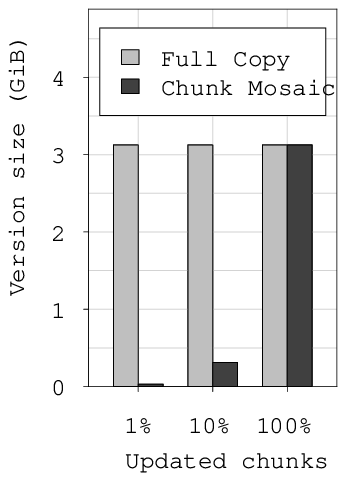}
\caption{New version size.}
\label{fig:exp:version:size}
\end{subfigure}
\begin{subfigure}[b]{1.6in}
\centering
\includegraphics[width=\columnwidth]{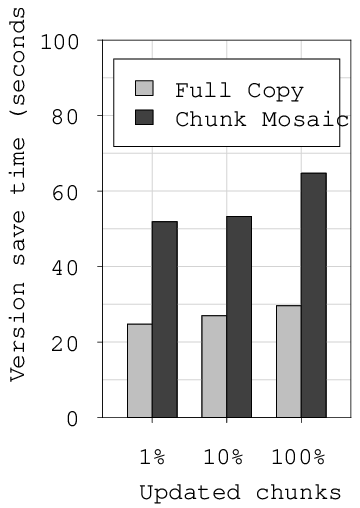}
\caption{Version save time.}
\label{fig:exp:version:file}
\end{subfigure}
\caption{Version size and save time to update a 3.1 GiB
dataset.}
\vspace{-1em}
\end{figure}

Figure \ref{fig:exp:version:size} compares the version size created by
the two time travel techniques. Writing a new version using Full Copy
occupies the same space as the
original one, since it always duplicates the previous dataset.
In contrast, the space usage of the Chunk Mosaic is proportional to the
number of chunks being updated. The Chunk Mosaic technique can thus
significantly reduce the storage footprint to support time travel
analysis.

Figure \ref{fig:exp:version:file} compares the time to save a new
version when the number of updated chunks varies from 1\% to
100\% of all chunks.
The Chunk Mosaic technique takes about
$2\times$ longer to write a version compared with Full Copy because it
reads and compares the data being saved with the previous version.
As the updated elements get dispersed to more chunks, the save time of
both techniques increases.


\vspace{-1em}
\section{Lessons Learned}
\label{sec:lessons}

\noindent
When designing \projectname,\ we learned the following lessons 
that are of broader interest to the designers of data management systems
for high-end computing facilities:

\textbf{1.~Reading directly from external array formats can be as fast as
reading from internal storage formats.}
We found that a carefully designed and tuned operator can read
dense external arrays in the HDF5 format as fast as reading from a native
SciDB array, which eliminates the processing and storage
overheads that are inherent in the data loading process.
This is due to the I/O-bound nature of many AQL/AFL queries and
also due to the fact that the HDF5 library does not introduce 
overheads that slow down disk I/O.
Scientific data format libraries like HDF5 are complex pieces of software with decades of
engineering and careful tuning based on actual deployment experience.
Although there is certainly potential for further improvement~\cite{tiledb}, 
existing array storage libraries should not be dismissed as
non-competitive without experimental evidence.

\textbf{2.~CPU overheads matter.}
In high-end computing facilities, CPU overheads can easily encumber the
I/O path and impact data processing throughput from cold storage.
Our experience contradicts a design tenet for database systems that
assumes an abundance of CPU cycles per I/O unit. 
In computers where the parallel file system offers massive I/O concurrency, 
this is only true when deploying at a scale of hundreds of nodes.
Because job scheduling in an HPC environment often gives higher priority
to small jobs that request a few nodes,
scientists prefer to run simple, time-sensitive analysis jobs on a
handful of nodes to reduce their job's queuing delay.

It is not uncommon to observe the I/O throughput from cold storage
reaching 3 GiB/sec per node, if only a few nodes are requested.
Such I/O rates require a careful investment of CPU cycles to
not make data processing CPU-bound.
\projectname\ eliminates substantial CPU overheads by masquerading the
HDF5 data as a special RLE chunk for processing (see
Algorithm~\ref{algo:scan_hdf5}) and enabling 
\emph{tiled mode} in SciDB which allows
more efficient chunk iteration by processing elements in a batch.
These optimizations alone improved the performance of the \texttt{scan\_hdf5()}
operator by more than $2\times$.

\textbf{3.~Disaggregating storage and compute requires dispensing
with static partitioning of data to instances.}
Shared-nothing database systems partition data on load under the
assumption of a fixed cluster size and hardware configuration. 
Systems designers need to dispense with the notion of static partitioning 
in a disaggregated datacenter that separates storage and compute:
in a scientific computer, users request some compute capacity and are
allocated to whatever nodes happened to be available first.
Therefore, the number of nodes the database system runs on changes between
jobs. This requires a flexible data partitioning policy that assigns
chunks to nodes at query time instead of during data loading.
In addition, the type of the compute node can change between jobs.
Hardware specifics like memory/CPU capabilities or MAC addresses should
ideally be enumerated at startup without requiring to recreate the database.

\textbf{4.~A non-materialized view mechanism is important for
interoperability.} 
The scientific computing software ecosystem is diverse and fragmented.
File format libraries are among the few software building
blocks that are widely used across scientific domains.
The file format library API mediates I/O accesses and is thus a natural
integration point to deliver more sophisticated data processing
capabilities to scientists. 
A view mechanism, such as virtual datasets in HDF5, decouples the
logical array content and physical data representation.
The view mechanism allows \projectname\ to break the SWMR constraint of
the HDF5 library and be backwards compatible with existing applications.
I/O middleware such as \projectname\ that support non-materialized views
can transparently convert between file formats and
improve the parallelization and update performance over the current HDF5
implementation.

\textbf{5.~The I/O hierarchy is complex in high-end computers, but caching is pervasive.}
When designing \projectname,\ we debated the appropriate integration
point to the SciDB engine. In particular, we contrasted the 
modularity of integrating at the query engine layer versus the
performance of optimizations such as buffer pool caching and
asynchronous I/O if one were to integrate at the storage layer.
\projectname\ integrates at the query engine layer.
Our evaluation reveals that repeated accesses on both cache-resident and
larger-than-memory datasets are cached by the parallel filesystem (see
Section~\ref{sec:exp:scan}). In retrospective, the performance benefit from explicitly
controlling I/O and caching would have been limited.


\section{Related work}
\label{sec:relwork}

\noindent\textbf{Array database systems} 

\noindent 
Database systems originally support
multi-dimensional datasets by materializing array offsets as attributes
of a relational table~\cite{baumann1994management}. 
Some solutions build new primitives to express array operations
within the DBMS~\cite{van2004ram,widmann1998efficient}, while
SciQL~\cite{zhang2011sciql} extends SQL to support the array data model.
Also, new query languages have been proposed for
querying arrays declaratively. AQL~\cite{libkin1996query} is one of the
earliest array query languages. AML~\cite{marathe2002query}
expresses array operations as a combination of operators, each of which
reads one or more arrays and returns an array.
Recent work has introduced techniques to
ameliorate skew during array join processing~\cite{duggan2015skew}
and has proposed similarity
joins for array data~\cite{zhao2016similarity}.

Another problem for array database systems is array storage. Of
particular research interest is storage of sparse datasets, as such
objects 
cannot be efficiently stored using a dense array representation. Soroush et al. investigate different
storage management and chunking strategies for array storage and
proposes a two-layer chunking strategy~\cite{soroush2011arraystore}.
TileDB proposes using \emph{flexible tiling} to address the
problem of accessing sparse data for efficient writes~\cite{tiledb}.
A related challenge is accessing
cells along different dimensions efficiently, to avoid
performing very small I/Os or reading unnecessary data. Novel chunking
and storage optimizations have been proposed to address this
problem~\cite{shimada2008storage}. 

SciDB~\cite{brown2010,stonebraker2009requirements} is an
open-source array database system. SciDB uses a similar chunk-based
storage as ArrayStore~\cite{soroush2011arraystore}, and
supports bitmap indices as well as RLE compression within a chunk to accelerate
sparse array processing.  It supports two query languages, an
operator-based language AFL (similar to AML), and a SQL-style query
language AQL.

\vspace{0.5em}
\noindent
\textbf{Scientific data formats} 

\noindent
HDF5~\cite{folk2011overview} and
NetCDF~\cite{rew1990netcdf} are two widely used portable scientific data
formats. The HDF5 library provides a set of tools to organize different
arrays and their metadata, such as particles~\cite{byna2013trillion} or
biological images~\cite{dougherty2009unifying} in a hierarchical,
semi-structured way. 
Optimizing HDF5 performance 
on HPC platforms is a well-studied problem; prior solutions have
explored how to tune large-scale array processing programs to improve
I/O performance \cite{byna2013trillion,howison2012tuning}.

\vspace{0.5em}
\noindent
\textbf{Array versioning} 

\noindent
Soroush et al.~propose versioned array storage to support both
querying a specific array version and extracting the history of an
array object~\cite{soroush2013time}.
Seering et al.~propose a materialization matrix to determine which version to
materialize while storing the remaining versions as deltas~\cite{seering2012efficient}. 
DataHub~\cite{bhardwaj2014datahub,bhardwaj2015collaborative}
extends the idea of version control to entire relational datasets for
data-intensive collaborations. Bhattacherjee et al.~investigate the
trade-offs between storage size and recreation time for versioning
techniques~\cite{bhattacherjee2015principles}.

\vspace{0.5em}
\noindent
\textbf{\emph{In situ} processing} 

\noindent
There is substantial prior research on querying external objects using a
relational database system.
NoDB~\cite{alagiannis2012nodb} proposed to cut the cost of data loading by
reading external files directly, and accelerated the querying process by
utilizing positional maps and selective parsing.
In a MapReduce context, invisible loading incrementally imports the data
into a database when parsing data in the map phase~\cite{abouzied2013invisible}.
The ScanRAW~\cite{cheng2015scanraw} operator queries the data \emph{in
situ}, while transparently loading frequent accessed data in the native storage of the
database.
ViDa moves beyond the \emph{in situ} data access model and proposes
just-in-time query processing~\cite{karpathiotakis2015just}.

Recent work has investigated \emph{in situ} mechanisms for data in the
HDF5 file format.
SDS/Q~\cite{blanas2014parallel, dong2014parallel} introduces a
light-weight relational query engine to directly process data in the
HDF5 format, in addition to using
bitmap-based indices~\cite{gosink2006hdf5,wu2005fastbit} to accelerate
highly-selective queries.
Wang
et al.~proposed running relational select and aggregation
queries directly on HDF5 file~\cite{wang2014saga,wang2013supporting}.
In the context of MapReduce, analysis frameworks have proposed querying HDF5 and
NetCDF data directly~\cite{geng2013scihive, wang2012scimate}.


\vspace{-1em}
\section{Conclusions}
\label{sec:conclusions}

We have presented ArrayBridge, a bi-directional array view mechanism for
scientific file formats that 
makes declarative array manipulations
interoperable with file-centric, imperative analysis kernels.
ArrayBridge bypasses the multi-hour load process of SciDB for massive
arrays.
Scientists can issue SciDB queries over multi-TiB external array datasets 
in the HDF5 format through ArrayBridge in a few minutes. ArrayBridge scales near-optimally
for cluster sizes up to 32 nodes until it encounters internal SciDB
scaling bottlenecks.
In addition to fast scans, ArrayBridge can materialize SciDB objects in
the HDF5 format nearly as efficiently as SciDB serializes its database
in its proprietary binary format for backup. 
For backwards compatibility,
existing applications access partitioned HDF5 objects as
a single logical dataset through a virtual view mechanism.
Finally, ArrayBridge can efficiently store versioned
objects by deduplicating unmodified chunks between versions. 
Past versions are accessible via the standard HDF5 API, which
enables time travel queries from version-oblivious imperative code.
The scan component of ArrayBridge is slated to be released in the 2017
community edition of SciDB.
We are deploying ArrayBridge in a large
scientific computing facility to make it readily available to scientists.

\section*{Acknowledgements}

This work was partly supported by the National Science Foundation 
under grants III-1422977, III-1464381, CNS-1513120,
by a Google Research Faculty Award, and by the DOE Office of Science, Advanced
Scientific Computing Research, under contract number DE-AC02-05CH11231.
The evaluation used resources of the National Energy Research Scientific
Computing Center (NERSC).

\Urlmuskip=0mu plus 1mu\relax

\scriptsize

\bibliographystyle{abbrv}
\vspace{1em}
\bibliography{paper}
\balancecolumns
\end{document}